\documentclass[10pt]{iopart}

\def\interi{\mathinner{\bf Z}}

\def\eequal{\equiv}

\def\poisson#1#2{\lbrace#1,#2\rbrace}
\def\fraz#1#2{{{#1}\over{#2}}}

\def\derpar#1#2{{{\partial#1}\over{\partial#2}}}

\def\Tau{{\cal T}}

\usepackage{epsfig}
\begin{document}

\title[Hannay angle]{The three-body problem and the Hannay angle}

\author{
Alessandro D.A.M. Spallicci\dag\ 
\footnote[3]{Corresponding author: UMR 6162, D\'ep. ARTEMIS d'Astrophysique Relativiste, BP 4229, Boulevard de l'Observatoire, 06304 Nice, France. Email: spallicci@obs-nice.fr}
Alessandro Morbidelli \dag\ Gilles Metris\dag\ }
\address{\dag\ Observatoire de la C\^ote d'Azur}

\begin{abstract}
The Hannay angle has been previously studied for a celestial circular restricted three-body system 
by means of an adiabatic approach.  
In the present work, three main results are obtained. 
Firstly, a formal connection between perturbation theory and the Hamiltonian adiabatic approach shows that 
both lead to the Hannay angle; it is thus emphasised that this effect is 
already contained in classical celestial mechanics, although not yet defined nor evaluated separately. Secondly,  
a more general expression of the Hannay angle, valid for an action-dependent  potential is given; such a 
generalised expression
takes into account that the restricted three-body problem is a time-dependent, two degrees of freedom 
 problem even when restricted to the circular motion of the test body. Consequently, (some of) the 
eccentricity terms cannot be neglected {\it a priori}. Thirdly, we present a new numerical estimate for the Earth adiabatically driven by Jupiter. We also point out errors in a previous  
derivation of the Hannay angle for the circular restricted three-body problem, with an action-independent  
potential.  

\end{abstract}


\begin{flushright}
{Interdisciplinary space science in Giuseppe Colombo's memory 
}
\end{flushright}

\section{Introduction}

During the 20 years since Berry \cite{ber84} discovered the quantum
geometric phase and Hannay \cite{han85} demonstrated its classical
counterpart, a great deal of research effort has gone into the subject of 
geometric phases. Shortly after their discovery, Berry \cite{ber85} showed how they 
were formally related. The Hannay effect is known to occur in a system undergoing 
an adiabatic change. The action variables behave as adiabatic
invariants, which remain unchanged as the system evolves. The conjugate angles,
however, being affected by the system's slow variation, exhibit a geometric shift which arises 
from the Hamiltonian being cyclic.    \\
From the mid 80s onwards, studies devoted to the Hannay angle, have analysed
several applications, including measurements in the solar system related to the
non-sphericity and slow rotation of the Earth \cite{celmech}. Instead, Berry
and Morgan \cite{bermor96} have attempted to determine the Hannay angle for
the restricted circular three-body problem. They give estimates for the
Earth and Mars revolving around the Sun, under the adiabatic influence of Jupiter, and
for a geostationary satellite, under the influence of the Moon.\\ 
We have structured the paper in the following manner. In the second section,
we derive the Hannay angle for the restricted circular three-body problem
based on classical perturbation analysis of celestial mechanics, using the same
simplified action-independent potential Hamiltonian of
\cite{bermor96}. This derivation simply underlines that the Hannay effect is not a new
{\it phenomenon} in celestial mechanics, but has always been taken into account
in all high order analytic theories. However, never before has it been defined nor
evaluated separately.
Furthermore, the previous derivation of the
Hannay angle \cite{bermor96} contains an error, (with reference to the masses of a three-body system, 
we let the attractor be A, the perturber P and the test mass be T)
 due to forcing T to move on a fixed, circular orbit. As a result, the interaction between P and A is 
neglected. 
In the third section, we start from a proper
representation of the restricted circular three-body problem, with an
action-dependent potential and derive a generalised Hannay angle.  
  We show that, although we are concerned by T circular motion solely,
  some eccentricity terms cannot be neglected {\it a priori}, as 
 the restricted three-body problem is intimately a time-dependent problem characterised by two degrees of freedom. 
 Neglecting all eccentricity terms as in \cite{bermor96}, implies an artificial 
  reduction to one degree of freedom (as if T were forced by an external constraint
  to move on a circular guide).
Finally, the appendix gives some useful basic expressions of the three-body problem.      

\section{The simplified restricted circular three-body problem}

\subsection{Adiabatic and perturbative approaches}

Berry and Morgan \cite{bermor96} have determined the Hannay angle for a 
simple system, inspired by the restricted circular three-body problem, with  
the following Hamiltonian: 

\begin{equation}
{\cal H}(q,p;t)=\frac{1}{2}p^2 + V_0 \cos(q-\epsilon t)\ ,
\label{Berry}
\end{equation}
where $V_0\sim {M_P}/{M_A}$ is a small parameter, of the order of the mass
of P relative to A, and
$\epsilon$ is another small parameter, proportional to the ratio of the orbital frequencies of P and T. 
The Hamiltonian (\ref{Berry}) is {\it not} the Hamiltonian of the restricted three-body problem, which we shall indeed consider in the next section. \\
Within the framework of the problem described by eq.(\ref{Berry}) and by the adiabatic theory, the Hannay angle 
was  approximated, at lowest order in $V_0$ by  \cite{bermor96}:

\begin{equation}
\theta_H=
2\pi \left (1 - \frac{\partial \nu}{\partial I} \right ) = 
-\frac{3\pi V_0^2}{I^4}\ ,
\label{eq:hannay}
\end{equation}
where $I$, the action, is the adiabatic invariant: 

\begin{equation}
I=\frac{1}{2\pi}\oint p{\rm d}q\ = 
 \frac{1}{2\pi} \int_0^{2\pi} \sqrt{2(E - V_0 \cos q)}dq 
\label{eq:ibm}
\end{equation}
$E$ is the T kinetic energy and $\nu = dE/dI$ is the angular
frequency. The Hannay effect is defined if closed curves of constant action
return to the same curves in phase space after a time evolution.  It is the
extra angle picked up by the angle variables, in addition to the time integral
of the instantaneous frequencies.  We remark that, given that the expression
of the Hannay angle in
(2) is negative, it causes a lengthening of the orbital period, conversely to
what is stated above eq.(49) in {\ref{Berry}.\\
We now re-derive the expression of the Hannay angle (\ref{eq:hannay}) for the
Hamiltonian (\ref{Berry}) but using the approach of classical perturbation
theory.  The first step is to re-write the Hamiltonian (\ref{Berry}) in
autonomous form. To this purpose, we introduce a pair of conjugate action
angle variables $\Tau, \tau\eequal \epsilon t$ and, in order to have
$\dot\tau=\epsilon$ in the Hamiltonian equations, write (\ref{Berry}) as:
\begin{equation}
{\cal H}(q,p,\tau,\Tau)=\frac{1}{2}p^2 + \epsilon \Tau + V_0 \cos(q-\tau)\ . 
\label{Berry-mio}
\end{equation}

We now consider $\frac{1}{2}p^2 + \epsilon \Tau$ as an unperturbed, integrable
Hamiltonian ${\cal H}_0$ and $V_0 \cos(q-\tau)$ as a small perturbation
${\cal H}_1$, of order 1 in $V_0$. Classical perturbation theory consists in
finding a change of variables $(q,p,\tau,\Tau)\to (q',p',\tau',\Tau')$ so as to remove
the term ${\cal H}_1$ from the new Hamiltonian (i.e. from the Hamiltonian
expressed in the new variables). The simplest way to achieve this result is to
introduce the new variables by Lie series (see ch. 2 of \cite{mor02}). Hence, we look for a transformation of variables of the
form 

\begin{equation} 
{p}=S_{\chi} p' \ ,\quad {q}=S_{\chi} q' \ ,\quad
{\Tau}=S_{\chi} \Tau' \ ,\quad {\tau}=S_{\chi} \tau' \ ,
\label{Lie-can-transf}
\end{equation} 
where $S_{\chi}$ is the Lie series operator, which, applied to a generic
function $f$, reads 

\begin{equation} S_{\chi}f= f+ \poisson{f}{\chi}+
\frac{1}{2}\poisson{\poisson{f}{\chi}}{\chi}+\ldots\ , 
\end{equation} 
where $\poisson{.}{.}$ denotes the Poisson bracket and $\chi$ is an {\it a priori}
unknown function.  The introduction of the new variables through
(\ref{Lie-can-transf}) has the advantage that (i) the new variables are
canonical {\it by construction}, and (ii) the Hamiltonian can be expressed in
the new variables in a straightforward manner, as

\begin{equation}
{\cal H}'=S_\chi {\cal H}={\cal H}+ \poisson{{\cal H}}{\chi}+
\frac{1}{2}\poisson{\poisson{{\cal H}}{\chi}}{\chi}+\ldots
\label{HtoH'}
\end{equation}
From the above expression, for the elimination of the term of order 1 in $V_0$ from ${\cal H}'$, 
the function $\chi$ must satisfy the equation

\begin{equation}
\poisson{{\cal H}_0}{\chi}+{\cal H}_1=0\ .
\label{homological}
\end{equation}
In our case, given the expression of ${\cal H}_0$ and ${\cal H}_1$, we need to set

\begin{equation}
\chi=-\frac{V_0}{p'-\epsilon} \sin(q'-\tau')\ .
\label{chi}
\end{equation}
Now that the function $\chi$ has been defined, terms of ${\cal H}'$ of 
order 2 in $V_0$ can be computed from (\ref{HtoH'}) and are:

\begin{equation}
\frac{1}{2}\poisson{\poisson{{\cal H}_0}{\chi}}{\chi}+\poisson{{\cal H}_1}{\chi}=
\frac{V_0^2}{2(p'-\epsilon)^2} \sin^2(q'-\tau')\ .
\label{quadratic}
\end{equation}
So, the new Hamiltonian up to order 2 in $V_0$ reads:

\begin{equation}
{\cal H}'=\frac{1}{2}p'^2 + \epsilon \Tau' + \frac{V_0^2}{2(p'-\epsilon)^2}
\sin^2(q'-\tau')\ .
\label{new-ham}
\end{equation}
We now average (\ref{new-ham}) with respect to $q'$. The averaged Hamiltonian
reads

\begin{equation}
\overline{{\cal H}}'=\frac{1}{2}p'^2 + \epsilon \Tau' +
\frac{V_0^2}{4(p'-\epsilon)^2}\ .
\end{equation}
The equation of motion for $q'$ is

\begin{equation}
\frac{\partial\overline{{\cal H}}'}{\partial p'} = {\dot q}'=p'-\frac{V_0^2}{2(p'-\epsilon)^3}\ .
\label{eq:caz}
\end{equation}
Referring to the slow periodic variation in time as modulation,  
this derivative can be split into an un-modulated frequency (evaluated for
$\epsilon=0$), namely $ p'-{V_0^2}/[2(p')^3]$, and a correction, obtained by derivation 
of (\ref{eq:caz})

\begin{equation}
\delta \dot{q}'= -3 \epsilon {V_0^2}/[2(p')^4]\ .
\label{eq:comecaz}
\end{equation} 

The Hannay angle is defined as the shift of $q'$ over a modulation period 
$2\pi/\epsilon$ with respect to the un-modulated case, and is therefore

\begin{equation}
\theta_H=\frac{2\pi}{\epsilon} \delta \dot{q}'= -\frac{3\pi V_0^2}{(p')^4}\ .
\label{hannay-mio}
\end{equation}

The expressions (\ref{eq:hannay}) and (\ref{hannay-mio}) are formally
identical, but the former is a function of $I$ while the latter is a function
of $p'$. We now show that $I$ and $p'$ coincide, up to order $O(V_0^2)$ and
$O(\epsilon)$. In fact, expanding (\ref{eq:ibm}) as a Taylor series of $V_0$ 
and recalling from (\ref{Berry}) that $E=p^2/2+V_0\cos(q-\epsilon t)$, we
obtain:
\begin{equation}
I=p+{{V_0}\over{p}}\cos(q-\epsilon t)+O(V_0^2)\ .
\label{I-expr}
\end{equation}
On the other hand, applying the transformation (\ref{Lie-can-transf}) with the
generating function $\chi$ given in  (\ref{chi}), we get:
\begin{equation}
p'=p+{{V_0}\over{p-\epsilon}}\cos(q-\tau)+O(V_0^2)\ .
\label{p'-expr}
\end{equation}
Therefore $p'\eequal I$ up to order 2 in $V_0$ and order 1 in $\epsilon$. 
This result, together with (\ref{eq:hannay}) and (\ref{hannay-mio})
proves that the adiabatic approach and the perturbation approach lead to 
an identical Hannay angle, at least for the 
Hamiltonian used in \cite{bermor96}.
 
\subsection{Hannay angle in the simplified problem}

The derivation in \cite{bermor96} forces the perturbed body to move on 
a circular orbit around the Sun, rather than considering from the beginning
the restricted three-body problem (RTBP). In this section we compute the
Hamiltonian that substitutes (1), if the RTBP is considered without constraints. 

As shown in the Appendix, the non Keplerian part of the potential in a three-body problem (to be added to the Keplerian AT term) is given
by\footnote{The potential in \cite{bermor96} has the dimensions of a frequency
  squared. Herein we use conventional units.}:

\begin{equation}
V \simeq G M_P \left (\frac{1}{R_{AP}}+  \frac{3}{2}\frac{R_{TA}^2}{R_{AP}^3} \cos^2\alpha -  
\frac{1}{2}\frac{R_{TA}^2}{R_{AP}^3}\right)
\label{eq:pot}
\end{equation}
where $R$ is the distance between bodies and $\alpha$ is the angle centred on A between T and P. We note that there is a  cancellation of the linear terms in
$\cos\alpha$, that was not previously considered, eq.(A4)
of \cite{bermor96}, see appendix\footnote {The source of 
the error in \cite{bermor96} is caused by forcing the Earth to move on a fixed curve (a 
circle), with two consequences. The former 
is the absence of the mutual attraction of P and A, second term of eq.(\ref{eq:potpart}), and thus the non-cancellation of the 
first order term due to the action of P on T, eq.(\ref{eq:yut}). As a result, the total action of P is 
proportional to $\frac{1}{R_{AP}}\left(\frac{R_{TA}}{R_{AP}}\right)^2$ 
instead of $\frac{1}{R_{AP}}\left(\frac{R_{TA}}{R_{AP}}\right)$ as in eq. 
(43) of \cite{bermor96}. The latter is to disregard {\it a priori} the potential terms 
proportional to the eccentricity, i.e. the removal of a degree 
of freedom. These terms induce a modification of the 
Hannay angle of order 0 in the eccentricity,  eq.(\ref{eq:14}) and do  
remain even if we set e=0 {\it a posteriori.}}\\
Since the term $1/R_{AP}$ does not depend on
T coordinates, the useful part of the perturbative function
reduces to:

\begin{equation}
    \label{eq:2}
V' =\frac{ G M_P} {R_{AP}}\left (\frac{R_{TA}}{R_{AP}}\right)^2 \frac{1}{4}
\left\{1+3\cos\left[2(\lambda_T-\lambda_P)\right]\right\}
\label{eq:simplepot}
\end{equation}
where we have used $\alpha=\lambda_T-\lambda_P$, $\lambda_T$ and
$\lambda_P$ being the true longitudes of $T$ and $P$
respectively.
The orbit of P is supposed to be circular; therefore
  $R_{AP}$ is constant and the true longitude $\lambda_P$ is equal to
  the mean longitude. But, as proved in section 3, for the orbit of T
  the circular approximation is not valid. 
  The perturbation needs to be expanded up to the first power of the
  eccentricity $e$ of T. Also, for an eccentric orbit the true
  longitude and the mean longitude are different, and it is more convenient
  to write the perturbation as a function of the latter. Using the
  classical expansions of the
  elliptic motion up to first power of the eccentricity
  \begin{eqnarray}
    \label{eq:1}
  R_{TA}&=&a\Big[1-e\cos(\lambda_{Tm}+\psi)\Big]\\
  \cos(2\lambda_T)&=&\cos(2\lambda_{Tm})+2e\left[\cos(3\lambda_{Tm}+\psi)-\cos(\lambda_{Tm}-\psi)
  \right]\\
  \sin(2\lambda_T)&=&\sin(2\lambda_{Tm})+2e\left[\sin(3\lambda_{Tm}+\psi)-\sin(\lambda_{Tm}-\psi) \right]
  \end{eqnarray}
we get:
\begin{equation}
  \label{eq:11}
  V'=\frac{ G M_P} {R_{AP}}\left (\frac{a}{R_{AP}}\right)^2 \frac{1}{4} 
\left\{1+3\cos2(\lambda_{Tm}-\lambda_P)+ \right. 
\end{equation}
{\mathindent=30pt
\[
   \left. e\left[-2\cos(\lambda_{Tm}+\psi)-9\cos(\lambda_{Tm}-\psi-2\lambda_P)+3\cos(3\lambda_{Tm}+\psi-2\lambda_P)\right]\right\}
\]
}
where $a$ is the semi-major axis of the orbit of $T$ around $A$, 
$\lambda_{Tm}$ is the mean longitude of $T$  and
$\psi$ is the opposite of the argument of its pericentre.

We now introduce the actions

\begin{equation}
  \label{eq:4}
  \Lambda_T=\sqrt{GM_A a}\qquad \Psi=\Lambda_T \left(1-\sqrt{1-e^2}\right)
\end{equation}
conjugate to the angle $\lambda_{Tm}$ and $\psi$ respectively. We make use of the third Keplerian
law relating the orbital frequency $\omega_T$ of T to the distance from A,

\begin{equation}
  \label{eq:7}
  \omega^2_T=\frac{GM_A}{a^3}
\end{equation}
Similarly we introduce the notation
\begin{equation}
  \label{eq:8}
  \omega_P^2=\frac{GM_A}{R_{AP}^3}
\end{equation}
and get

\begin{equation}
  \label{eq:5}
 V' =\omega_P^2\frac{M_P} {M_A}\frac{\Lambda_T}{\omega_T} \frac{1}{4}
\left\{(1+3\cos\left[2(\lambda_{Tm}-\lambda_P)\right]+ \right.\\
\end{equation} 
{\mathindent=10pt
\[
\left.\sqrt{\frac{2\Psi}{\Lambda_T}}\left[-2\cos(\lambda_{Tm}+\psi)-9\cos(\lambda_{Tm}-\psi -2\lambda_P)+3\cos(3\lambda_{Tm}+\psi -2\lambda_P)\right]\right\}\nonumber
\]
}
Now, $V'$ depends on the actions and the formalism  used by Berry and Morgan is no longer
directly applicable. 
Instead, the Hannay angle will be derived by
means of a
perturbation approach in the next section.\\
The expression (\ref{eq:2}) of the perturbation is derived from an
expansion of $1/R_{AP}$ in powers of $R_{TA}/R_{AP}$ limited to the second
  order. This expansion by means of Legendre polynomials is
  inaccurate when  $R_{TA}/R_{AP}$ is not small. In this case,
    a Fourier like expansion, using Laplace coefficients \cite{lap},
 is more appropriate:

    \begin{equation}
      \label{eq:3}
      \frac{R_{TA}}{R_{AP}}\simeq \frac{a}{R_{AP}}\left[ 1-\sqrt{\frac{2\Psi}{\Lambda_T}}\cos(\lambda_{Tm}+\psi)\right ]
=
\end{equation}
{\mathindent=40pt 
\[     
 \left[\frac{1}{2}b^{(0)}_{1/2}+\sum_{k=1}^{\infty}b^{(k)}_{1/2}\cos(k\alpha)\right]\left[1-\sqrt{\frac{2\Psi}{\Lambda_T}}\cos(\lambda_{Tm}+\psi)\right ]
\]
}    
where the Laplace coefficients $b^{(k)}_{1/2}$ are functions of
    $\displaystyle \frac{a}{R_{AP}}=\frac{\Lambda^2_T}{GM_AR_{AP}}$. Finally, the potential takes the form:

\begin{equation}
  \label{eq:6}
  V'' = \frac{M_P} {M_A}\sum_{k} B_k(\Lambda_T)\cos [k(\lambda_{Tm}-
  \lambda_P)] +
\label{eq:potlap}
\end{equation}
{\mathindent=0pt
\[  
C_k^+(\Lambda_T)\sqrt{\Psi}\cos[k\lambda_{Tm}-\psi - (k+1)\lambda_P]  + 
C_k^-(\Lambda_T)\sqrt{\Psi}\cos[k\lambda_{Tm}+\psi - (k-1)\lambda_P]
\]
} 

\section{The Hannay angle in the restricted circular three-body problem}

\subsection{Correct derivation of the Hannay angle}

In this section we derive the expression of the Hannay angle, starting from
the correct representation of the RTBP (see sect. 2.2).
Because the RTBP has three degrees of freedom, the use of the adiabatic
  approach is rather cumbersome. Therefore we prefer to use a perturbative
  approach. As shown in the example of the previous section, with a
  perturbation approach we can easily isolate the term that corrects the
  orbital frequency of the small body, at first order in the slow frequency
  $\epsilon$ of P.  This term, evaluated in
  the following, is the time derivative of the Hannay angle.
  
  Although we are ultimately interested in evaluating the Hannay angle for a
  circular motion of T, the eccentricity terms in the perturbation
  function of the RTBP cannot be neglected {\it a priori}. Indeed, in section
  2.1 we have seen that the Hannay angle arises from the quadratic term in P mass, which is generated from the Poisson bracket between the
  original perturbation and the generating function. In the Poisson bracket
  operation, linear terms in the eccentricity (i.e. proportional to
  $\sqrt{\Psi}$) can produce eccentricity independent terms.
  On the other hand, terms of higher order in the eccentricity cannot produce
  such terms. Therefore, it is necessary, and sufficient, to consider the
  expansion of the original Hamiltonian up to order 1 in $\sqrt{\Psi}$ only,
  namely:

\begin{equation}
{\cal H}=-\frac{1}{2\Lambda^2_T} + \epsilon \Lambda_P + m_P \sum_{k\in \interi} \left\{b_k(\Lambda_T)
\exp[\iota k(\lambda_{Tm}-\lambda_{Pm})] + \right.
\label{ciccio}
\end{equation}
{\mathindent=0pt
\[
\left.
c_k^+(\Lambda_T)\sqrt{\Psi}\exp\{\iota[k\lambda_{Tm}-(k+1)\lambda_{Pm}-\psi]\}
+
c_k^-(\Lambda_T)\sqrt{\Psi}\exp\{\iota[k\lambda_{Tm}-(k-1)\lambda_{Pm}+\psi]\}\right\}
\]}
where $\lambda_{P}$ is the mean longitude of P, $\Lambda_P$ is the
conjugate action of P (identical to true longitude for a circular orbit), $m_P$ is the mass of P 
relative to A and $\epsilon$ is the orbital frequency of the
P relative to T. In (\ref{ciccio}) we have assumed
  $G=M_A=1$ and omitted these quantities for simplicity of notation.  The
symbol $\iota$ denotes the imaginary unit ($\sqrt{-1}$). For d'Alembert rules,
the coefficients $b_k, c_k^+, c_k^-$ are real and $b_k=b_{-k}$,
$c_k^+=c_{-k}^-$ (see ch. 1 of \cite{mor02}). In (\ref{ciccio}) the motion of
$\lambda_P$ is assumed to be linear in
time, which reflects the linear dependence of ${\cal H}$ on $\Lambda_P$.
As in the previous section, terms independent of the angles $\lambda_{Tm},
\lambda_P, \psi$ are grouped in the integrable approximation ${\cal H}_0$ and
the harmonic terms constitute the perturbation ${\cal H}_1$.

Using (\ref{HtoH'}) and the new expressions for ${\cal H}_0$ and ${\cal H}_1$, 
the function $\chi$ required to eliminate ${\cal H}_1$ from the new Hamiltonian is:

{\mathindent=0pt
\begin{equation}
\chi = m_P\sum_{k\in\interi\setminus 0}\left\{ \frac{\iota
  b_k(\Lambda)}{\left({\displaystyle \frac{1}{\Lambda^3_T}}-\epsilon\right)k} \exp[\iota
  k(\lambda_T -\lambda_P)]+\right.
\end{equation}}
\[
\frac{\iota
  c_k^+(\Lambda_T)\sqrt{\Psi}}{\left[{\displaystyle \fraz{k}{\Lambda^3_{Tm}}}-
(k+1)\epsilon\right]}\exp\{\iota[k\lambda_{Tm}-(k+1)\lambda_P-\psi]\} +
\]
\[
\left.
\frac{\iota  c_k^-(\Lambda_T)\sqrt{\Psi}}{\left[{\displaystyle\fraz{k}{\Lambda^3_{Tm}}}-(k-1)\epsilon\right]}
\exp\{\iota[k\lambda_{Tm}-(k-1)\lambda_P+ \psi]\}\right\} ,
\] 
where for simplicity we have omitted the primed notation for the new variables.
This function is well defined if $1/\Lambda^3_T\ne \epsilon$, and if 
$1/\Lambda^3_T\ne  \epsilon (k\pm{}1)$, which is true away from 
the 1/1 mean motion resonance and from all mean motion resonances of first
order in the eccentricity. 

The terms of the new Hamiltonian that are of order 2 in $M_P$ can be
computed by the left hand side of the expression (\ref{quadratic}). 
 After that the Poisson brackets are calculated in (\ref{quadratic}), we
  can finally neglect the resulting terms that depend on the eccentricity, and
  retain only those of order 0 in $e$ (i.e. the terms independent of
  $\Psi$), which are:

\begin{equation}
{\cal H}'_2=\frac{M_P^2}{2}\sum_{k}\sum_{l\ne 0}d_{k,l}\exp[\iota
(k+l)(\lambda_{Tm}-\lambda_{P})]
\label{H2}
\end{equation}
with

{\mathindent=0pt
\begin{equation}
d_{k,l}=  
-\derpar{b_k}{\Lambda_T}\frac{b_l}{\left({\displaystyle\frac{1}{\Lambda^3_T}}-\epsilon\right)} 
+ k b_k\left [\derpar{b_l}{\Lambda_T}\frac{1}{\left({\displaystyle
        \frac{1}{\Lambda^3_T}}-\epsilon\right)l}+\frac{3b_l}{\Lambda^4_T\left({\displaystyle\frac{1}{\Lambda^3_T}}-\epsilon\right)^2l}\right ]\
\end{equation}}
\[
-\fraz{c_k^+c_l^-}{\left[{\displaystyle\fraz{l}{\Lambda^3_T}}-(l-1)\epsilon\right]}+
\fraz{c_k^- c_l^+}{\left[{\displaystyle\fraz{l}{\Lambda^3_T}}-(l+1)\epsilon\right]}
\label{coef}
\]

As in the previous section, we now average the new Hamiltonian over
$\lambda_{Tm}$. Of all the terms in (\ref{H2}) only those with $k=-l$ survive, 
because they are independent of $\lambda_{Tm}$. The expansion of these terms at
order 1 in $\epsilon$ gives the function

{\mathindent=0pt
\begin{equation}
{\cal F}=\epsilon\frac{m^2_P}{2}\sum_{k\ne 0} 
\left\{\Lambda^6_T \left[-2 b_k \derpar{b_k}{\Lambda_T}
    +(c_k^+)^2\fraz{(k+1)}{k^2}-(c_k^-)^2\fraz{(k-1)}{k^2}\right] - 6
    \Lambda^5_T b_k^2 \right\}\ .
\label{final-1}
\end{equation}}

The change in the frequency of $\lambda_{Tm}$ due to the slow motion of $\lambda_P$ 
is then
\begin{equation}
\delta\dot\lambda_{Tm}=\fraz{{\rm d}{\cal F}}{{\rm d}\Lambda_T}
\end{equation} 
and the Hannay angle becomes
\begin{equation}
\theta_H=\frac{2\pi}{\epsilon} \delta\dot\lambda_{Tm}\ .
\label{final}
\end{equation}

\subsection{Correct expression for the Hannay angle}

In order to apply the results of the previous section  to the
perturbation (\ref{eq:5}), limited to the power 2 of $R_{TA}/R_{AP}$, we set

\begin{equation}
  \label{eq:9}
  b_2=b_{-2}=-\frac{3}{8}\frac{\omega_P^2}{M_A}\frac{\Lambda_T}{\omega_T}
\end{equation}
for the terms independent of the
eccentricity, and 

\begin{eqnarray}
  \label{eq:13}
 c_1^-=c_{-1}^+&=&\frac{1}{4}\frac{\omega_P^2}{M_A}\frac{\sqrt{2\Lambda_T}}{\omega_T}\\
 c_1^+=c_{-1}^-&=&\frac{9}{8}\frac{\omega_P^2}{M_A}\frac{\sqrt{2\Lambda_T}}{\omega_T}\\
 c_3^-=c_{-3}^+&=&-\frac{3}{8}\frac{\omega_P^2}{M_A}\frac{\sqrt{2\Lambda_T}}{\omega_T}
\end{eqnarray}
for the terms proportional to the eccentricity.
Using (\ref{final-1})--(\ref{final}) we get the following
contributions to the Hannay angle:
\begin{equation}
  \label{eq:10}
  \theta_H^{(c)}=-\frac{819}{16}\pi\left(\frac{M_P}{M_A} \right)^2\left(\frac{a}{R_{AP}} \right)^6
\end{equation}
for the terms not related to the eccentricity and 

\begin{equation}
  \label{eq:14}
 \theta_H^{(e)}=130\pi\left(\frac{M_P}{M_A} \right)^2\left(\frac{a}{R_{AP}} \right)^6
\end{equation}
for the terms coming from the eccentricity. Note that the major
contribution of the last term which changes the
sign of the effect. In \cite{bermor96}, there are two errors concerning the sign. Firstly, the sign of the total effect is indeed positive and not negative. Secondly, the negative sign would be equivalent to a lengthening of the orbit and not to shortening as stated before eq.49 in \cite{bermor96}.

Finally, the total Hannay angle due to P is given by:
\begin{equation}
  \label{eq:15}
\theta_H=\frac{1261}{16}\pi\left(\frac{M_P}{M_A} \right)^2\left(\frac{a}{R_{AP}} \right)^6  
\end{equation}

The Hannay angle of the Earth due to the perturbation by Jupiter
amounts to $1.14 \times{}10^{-8}$ radians which is 1705 m per Jupiter orbit, i.e. 144 m per year, of forward displacement, or 4.8 ms of orbital period shortening per year. Furthermore, 
the value of the Hannay angle is comparable (apart the sign) to the value found in \cite{bermor96} even if the ratio $R_{TA}/R_{AP}$ is to the sixth power rather than fourth as in \cite{bermor96}, since there is 
A large numerical factor $(1261/16)$, eq.\ref{eq:15}, instead of $3$ in \cite{bermor96}.\\
The Hannay effect on an artificial satellite adiabatically driven by the Moon as indicated by \cite{bermor96} is more complex, especially if coupled to the issue of measurability, and it deserves a separate publication.  
  
\section{Conclusions}
The relation of the Hannay angle to classical perturbation theory of celestial
mechanics has been {\it given}, for both action-dependent and independent perturbations. The expression of the Hannay angle in a restricted circular three-body problem has been properly derived. 

\section{Acknowledgments} 
 Dr S. Feltham (ESA), Dr Ch. Salomon (Ecole Normale Sup. Paris) and Prof. Sir M. Berry, Bristol Univ., are acknowledged for various discussions. The European Space Agency is acknowledged for awarding the Senior Research Fellowship G. Colombo to A. Spallicci.

\section*{Appendix for the three-body problem}

\begin{figure}[tbp]
\begin{center}
\epsfig{file=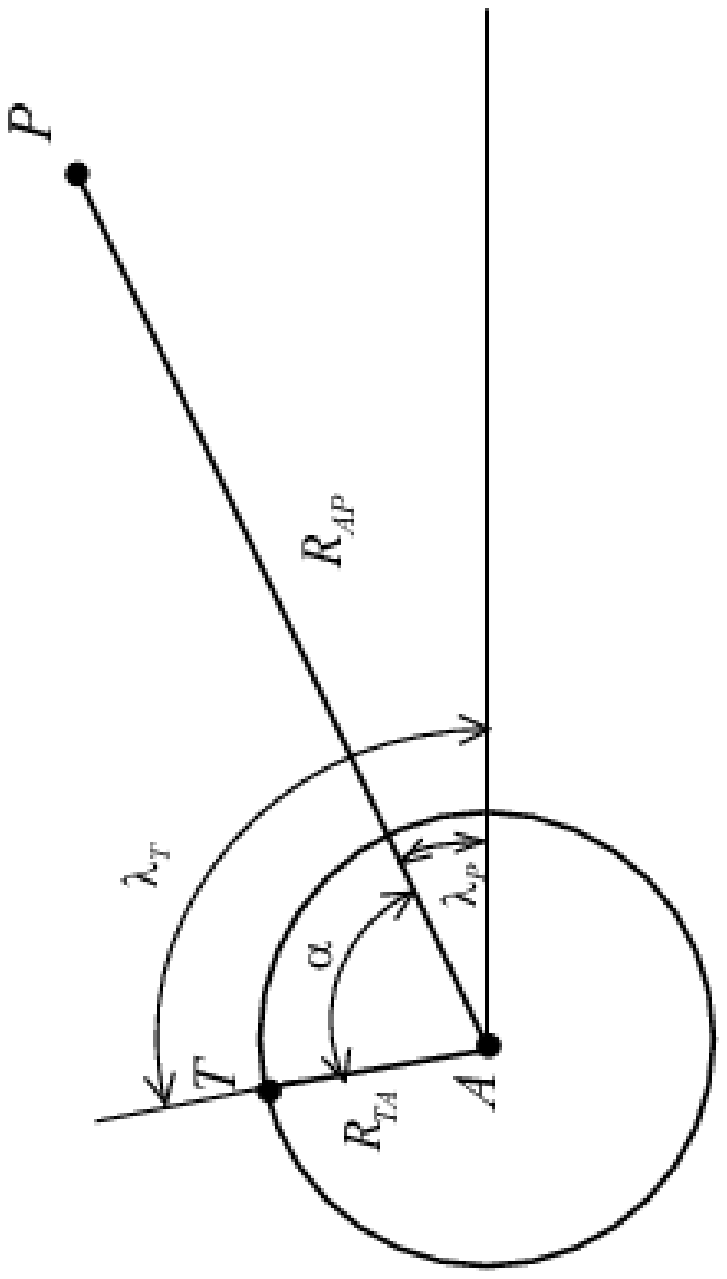}
\end{center}
\caption{The restricted three-body problem main variables. }
\label{fig:fig. 1}
\end{figure}

The total potential energy of the system is \cite{plu}:

\begin{equation} 
E_V = - \left( \frac{M_A M_P}{R_{AP}} + \frac{M_T M_P}{R_{TP}} + \frac{M_T M_A}{R_{TA}}\right ) G 
\label{eq:vplu}
\end{equation}

where $R$ is the interdistance among bodies:

\begin{equation}
R_{AP} = \left[(x_A - x_P)^2 + (y_A - y_P)^2 + (z_A - z_P)^2 \right ]^{1/2}
\end{equation}

\begin{equation}
R_{TP} = \left[(x_T - x_P)^2 + (y_T - y_P)^2 + (z_T - z_P)^2 \right ]^{1/2}
\end{equation}

\begin{equation}
R_{TA} = \left[(x_T - x_A)^2 + (y_T - y_A)^2 + (z_T - z_A)^2 \right ]^{1/2}
\end{equation}

The equations of motion of A and T are, respectively:

\begin{equation}
M_T{\ddot x}_T = - \frac{\partial E_V}{\partial x_T}~~~~~~~~M_T{\ddot y}_T = - \frac{\partial E_V}{\partial y_T}~~~~~~~~M_T{\ddot z}_T = - \frac{\partial E_V}{\partial z_T}
\end{equation}

\begin{equation}
M_A{\ddot x}_A = - \frac{\partial E_V}{\partial x_A}~~~~~~~~M_A{\ddot y}_A = - \frac{\partial E_V}{\partial y_A}~~~~~~~~M_A{\ddot z}_A = - \frac{\partial E_V}{\partial z_A}
\end{equation}

For $(\xi, \eta, \zeta)$ relative coordinates of T and P, relative to A: 

\[
x_T = x_A + \xi_T~~~~~~~~~~y_T = y_A + \eta_T~~~~~~~~~~z_T = z_A + \zeta_T 
\]

\[
x_P = x_A + \xi_P~~~~~~~~~~y_P = y_A + \eta_P~~~~~~~~~~z_P = z_A + \zeta_P
\]
we have:
\[
{\ddot \xi}_T = - \frac{1}{M_T}\frac{\partial E_V}{\partial x_T} + \frac{1}{M_A}\frac{\partial E_V}{\partial x_A} =
\]
{\mathindent=20pt
\[
\left [
-  \frac{M_P(x_T - x_P)}{R_{TP}^3} - \frac{M_A(x_T - x_A)}{R_{TA}^3} + \frac{M_P(x_A - x_P)}{R_{AP}^3} +
\frac{M_T(x_A - x_T)}{R_{TA}^3} 
\right ]G =
\]}
\begin{equation}
\left [
-  \frac{(M_T + M_A)\xi_T}{R_{TA}^3} - \frac{M_P(\xi_T - \xi_P)}{R_{TP}^3} - \frac{M_P\xi_P}{R_{AP}^3} 
\right ]G
\end{equation}

Setting 

\begin{equation}
V = \left [
\frac{M_P}{R_{TP}} - \frac{M_P}{R_{AP}^3}(\xi_T\xi_P + \eta_T\eta_P + \zeta_T\zeta_P) 
\right ]G
\end{equation}
we finally have:

\begin{equation}
{\ddot \xi}_T = - (M_T + M_A)G\frac{\xi_T}{R_{TA}^3} + \frac{\partial V}{\partial \xi_T}
\end{equation}

\begin{equation}
{\ddot \eta}_T = - (M_T + M_A)G\frac{\eta_T}{R_{TA}^3} + \frac{\partial V}{\partial \eta_T}
\end{equation}

\begin{equation}
{\ddot \zeta}_T = - (M_T + M_A)G\frac{\zeta_T}{R_{TA}^3} + \frac{\partial V}{\partial \zeta_T}
\end{equation}

For $\alpha$ angle between T and A:

\begin{equation}
V = G M_P \left (
\frac{1}{R_{TP}} - \frac{R_{TA}}{R_{AP}^2}\cos \alpha 
\right )
\label{eq:potpart}
\end{equation}
and using Legendre polynomials $P_n(x)$, the inverse of the distance between P and {T} is written as: 

\begin{equation}
\frac{1}{R_{TP}} = \frac{1}{\left(R_{TA}^2 + R_{AP}^2 - 2 R_{TA}R_{AP}\cos\alpha \right )^{1/2}} =
\label{eq:yut}
\end{equation}
{\mathindent=0pt
\[
\frac{1}{R_{AP}}\sum_{n=0}^\infty P_n(\cos\alpha) \left (\frac{R_{TA}}{R_{AP}}\right)^n \simeq
\frac{1}{R_{AP}}\left[ 1 +  \frac{R_{TA}}{R_{AP}} \cos\alpha + \frac{1}{2} (3 \cos^2 \alpha - 1 )
\left (\frac{R_{TA}}{R_{AP}}\right)^2\right] 
\]}
Substituting eq.(\ref{eq:yut}) into eq.(\ref{eq:potpart}), gives rise to eq.(\ref{eq:pot}).

\end{document}